\documentclass[prb,floats,twocolumn]{revtex4}
\usepackage{graphics,dcolumn}
\begin{document}

\title{Towards an assessment of the accuracy of density functional
theory for first principles simulations of water II }
\author{Eric Schwegler}
\author{Jeffrey C. Grossman}
\author{Fran\c{c}ois Gygi}
\author{Giulia Galli}
\affiliation{Lawrence Livermore National Laboratory,
P.O. Box 808, Livermore, California 84559}

\begin{abstract}
A series of 20 ps {\it ab initio} molecular dynamics simulations of
water at ambient density and temperatures ranging from 300 to 450~K
are presented.  Car-Parrinello (CP) and Born-Oppenheimer (BO)
molecular dynamics techniques are compared for systems containing 54
and 64 water molecules. At 300~K, excellent agreement is found between
radial distribution functions (RDFs) obtained with BO and CP dynamics,
provided an appropriately small value of the fictitious mass parameter
is used in the CP simulation. However, we find that the diffusion
coefficients computed from CP dynamics are approximately two times
larger than those obtained with BO simulations for T$>$400~K, where
statistically meaningful comparisons can be made.  Overall, both BO
and CP dynamics at 300 K yield overstructured RDFs and slow diffusion
as compared to experiment. In order to understand these discrepancies,
the effect of proton quantum motion is investigated with the use of
empirical interaction potentials. We find that proton
quantum effects may have a larger impact than previously thought on
structure and diffusion of the liquid.
\end{abstract}

\date{\today}
\maketitle

\section{Introduction}

Recently, a number of studies have focused on understanding in detail
the level of accuracy that can be achieved in {\it ab initio}
molecular dynamics simulations of water
\cite{sizvekov02,dasthagiri03,jgrossman04}.  In particular, it was
found in both Ref.~\cite{dasthagiri03} and \cite{jgrossman04} that in
the absence of proton quantum effects, density functional theory (DFT)
within widely used generalized gradient approximations (PBE
\cite{jperdew96} and BLYP \cite{abecke88,clee88}) leads to
significantly overstructured radial distribution functions (RDFs) and
slower diffusion than experiment, at ambient temperature and pressure.

In our previous work \cite{jgrossman04}, within the Car-Parrinello (CP) 
method \cite{rcar85}, we found
that using either the Perdew-Burke-Ernzerhof (PBE)
\cite{jperdew96} or the Becke-Lee-Yang-Parr (BLYP)
\cite{abecke88,clee88} functional leads to 
very small differences in the structural properties of the room
temperature liquid. Size effects, although not fully
negligible when using 32 molecule cells, were found to be rather
small.  We also found that there is a wide range of ratios ($\mu/M$)
between the fictitious electronic mass ($\mu$) and the smallest ionic
mass of the system (either $M\sim1836$ au for hydrogen or $M\sim3672$ au
for deuterium atoms) for which the electronic ground state is accurately 
described in microcanonical CP simulations ($\mu/M$ $\leq$ 1/5). However, 
care must be exercised not to carry out simulations outside this range, 
where structural properties may artificially depend on $\mu$.  Specifically,
our results showed that the use of $\mu/M \sim1/3$ leads
to an artificial softening of the pair correlation functions, which are in
fortuitous agreement with experiment. In the case of an accurate
description of the electronic ground state, and in the absence of
proton quantum effects, the oxygen-oxygen RDF is found to be
over-structured compared with that obtained in neutron scattering 
\cite{asoper00b} and x-ray diffraction experiments \cite{jsorenson00} 
and the computed diffusion coefficients are 10 to 20 times smaller 
than experiment.  In
our previous work \cite{jgrossman04}, we also examined the 
amount of sampling  needed to
obtain statistically independent data points for calculated properties,
and showed that for segments of the oxygen-oxygen RDF the data correlation 
time is quite long; time intervals
as large as $\sim$10-20 ps are required to obtain a single statistically 
independent data point (correlation time is not to
be confused with equilibration time, as we will discuss later in the
text.)

Although a number of important technical parameters were tested in
Ref. \cite{jgrossman04}, several questions remain unanswered. First,
it is important to fully test the convergence of the CP algorithm as
the value of the fictitious mass approaches zero and to investigate
whether possible sources of inaccuracies, such as those pointed out by
Tangney and Scandolo \cite{ptangney02b} persist even for small $\mu/M$
ratios ($\mu/M$ $\leq$ 1/5). In order to address this issue, we
present here a series of {\it ab initio} MD simulations at different
temperatures, using both the CP method and minimizing the Kohn-Sham
energy functional at each MD time step, while keeping all the other
parameters of the two calculations fixed. We refer to this direct
minimization technique as Born-Oppenheimer (BO) MD. Our results
show an excellent agreement between RDFs obtained in CP and BO
simulations, provided an appropriately small value of the fictitious
electronic mass is used ($\mu/M$ $\leq$ 1/5), in agreement with
previous BO simulations \cite{dasthagiri03}. However, diffusion 
coefficients obtained within BO and CP
simulations appear to differ by about a factor of two at temperatures
(T) of $\simeq$ 400K. At this temperature, a statistically meaningful
comparison is possible using the simulation times 
employed here (about 20 ps). Unfortunately, at lower temperatures such a
quantitative comparison will require longer simulation times. 

In addition to comparing BO and CP simulations, we have carried out a
detailed analysis of the effect of proton quantum motion on computed
structural and dynamical properties, by using the results of extended
classical simulations with empirical inter-atomic forces. Our findings
indicate that the inclusion of proton quantum effects would bring the
DFT results obtained here in much better agreement with experiment,
for both RDFs and diffusion coefficients, and that their inclusion is
crucial to accurately describe the properties of water at ambient
conditions. In agreement with most previous studies, we find that
proton quantum effects tend to give softer RDFs and faster diffusion
than classical (Newtonian) simulations. We also find that the effect
of proton quantum motion on calculated RDFs and diffusion coefficients
at 300 K is the same as that obtained by increasing the ionic temperature by
about 50 or 100~K at constant density, when using a rigid or 
flexible water model, respectively.

The rest of the paper is organized as follows. Section \ref{methods} contains a
description of the methods used in both {\it ab initio} and classical MD
simulations of water and Section \ref{results} describes our results in
detail. In particular, in Section \ref{BOvsCP} we report a comparison between CP and
BO simulations, and in Section \ref{errors} we discuss possible sources of the
apparent disagreement with experiment, namely the absence of proton
quantum effects and the GGA functional (PBE) adopted here. The effect
of temperature variations between 300 and 450~K on water at ambient
density is analyzed in Section \ref{temp}. Finally, Section \ref{conc} contains our
conclusions.

\section{Methods}\label{methods}

We have performed a series of molecular dynamics simulations \cite{qbox} of liquid
water with both {\it ab initio} and classical potentials.  The {\it ab
initio} molecular dynamics simulations are based on density functional
theory with the PBE \cite{jperdew96} generalized gradient
approximation. Electron-ion interactions are treated with norm
conserving non-local pseudopotentials of the Hamann type
\cite{dhamann89,lkleinman82}, and with the Kohn-Sham orbitals and
charge density expanded in plane waves up to a kinetic energy cutoff
of 85 and 340 Ry, respectively. In each simulation, the starting
configuration was generated by performing 500 ps of classical
molecular dynamics with the TIP5P potential \cite{mmahoney00},
followed by 3 ps of {\it ab initio} molecular dynamics with a weakly
coupled velocity scaling thermostat set at the given target
temperature. The thermostat was then removed and all reported
statistics were collected under constant energy conditions.

For the Born-Oppenheimer (BO) simulations, the system consists of 64
water molecules in a cubic box of length 12.43 \AA\ with periodic
boundary conditions.  At each molecular dynamics timestep, the
wavefunction was relaxed to the ground state by preconditioned
steepest descent with Anderson acceleration \cite{danderson65}. For a
timestep of 0.24 fs, 12 electronic iterations were found to be
sufficient to converge the total energy to within $10^{-9}$ a.u. at
each molecular dynamics iteration. Under these simulations conditions,
the drift in the total energy was found to be equivalent to a 0.27
K/ps increase in the system temperature.

In the Car-Parrinello (CP) simulations, a slightly smaller system size
of 54 water molecules in a cubic box of length 11.7416 \AA\ was
used. The electronic degrees of freedom were propagated with a
fictitious mass of $\mu$=340 a.u., and with a molecular dynamics
timestep of 0.07 fs.

\begin{table}[t]
\caption{Geometry and potential parameters defined as in Refs.
\cite{mmahoney00,mmahoney01} for the TIP5P water
models.}
\begin{ruledtabular}
\begin{tabular}{lccc}
                             &  TIP5P  & TIP5P(PIMC) \\ \tableline
q$_H (e)$                    &  0.241  &  0.251      \\
q$_L (e)$                    & -0.241  & -0.251      \\
$\sigma_O$ (\AA)             &  3.12   &  3.12       \\
$\epsilon_O$ (kcal/mol)      &  0.16   &  0.16       \\
$r_{OH}$ (\AA)               &  0.9572 &  0.9572     \\
$r_{OL}$ (\AA)               &  0.70   &  0.70       \\
$\theta_{HOH}$ (deg)         &  104.52 &  104.52     \\
$\theta_{LOL}$ (deg)         &  109.47 &  109.47     \\
\end{tabular}
\end{ruledtabular}
\label{watermodels}
\end{table} 

All classical simulations were performed with the Tinker simulation
package \cite{jwponder03} under constant volume and temperature 
conditions for systems of 64 rigid water molecules. A molecular dynamics 
timestep of 0.5 fs was used and the simulation temperature was maintained 
with a weakly coupled Berendsen-type thermostat \cite{hberendsen84}. 
Long-range electrostatic interactions were included with the particle-mesh
Ewald method \cite{tdarden93}. Both the TIP5P
\cite{mmahoney00} and the TIP5P(PIMC) \cite{mmahoney01}
potentials, which are defined in Table \ref{watermodels}, have been
examined.

\section{Results}\label{results}

\subsection{Comparison of CP and BO dynamics at 300 K}\label{BOvsCP}

In Ref.~\cite{jgrossman04}, we have shown that in CP simulations of
water at ambient conditions, one can vary the fictitious mass
parameter, $\mu$, within a suitable range without significantly
altering the results of the simulations. In particular, it was found
that when $\mu/M\ \le 1/5$, where $M$ is the mass of the H or D atom
in the simulation, the average structural properties and computed
diffusion coefficients appear to be insensitive to the choice of
$\mu$. This is in contrast to simulations performed with large values
of $\mu$ that lead to significant changes in both the structure and
diffusion of the liquid. As discussed in Ref.~\cite{jgrossman04}, the
reason for the dramatic changes observed as $\mu/M$ is increased from
$1/5$ are due to a direct overlap of the fictitious electronic spectra
with the highest frequency ionic spectra. In other words, as long as
the fictitious electronic degrees of freedom can be adiabatically
decoupled from the ionic degrees of freedom, the Car-Parrinello
technique yields accurate results. However, as argued by Tangney {\it
et al.}  \cite{ptangney02b}, exact adiabatic separation in a
Car-Parrinello simulation is never actually achieved, even for
relatively small values of $\mu$. The reason for this is because in
addition to the high frequency components in the ionic spectra that
can lead to direct coupling to fictitious electronic degrees of
freedom, there are additional $\mu$-dependent errors associated with
the low-frequency components of the ionic motion. The impact of a
particular choice of $\mu$ can be assessed in a direct fashion by
comparing Car-Parrinello simulations to so-called Born-Oppenheimer
(BO) simulations of water, where the wavefunctions are converged to
the ground state at each molecular dynamics time step.

\begin{figure}
\rotatebox{-90}{\resizebox{2.5in}{!}{\includegraphics{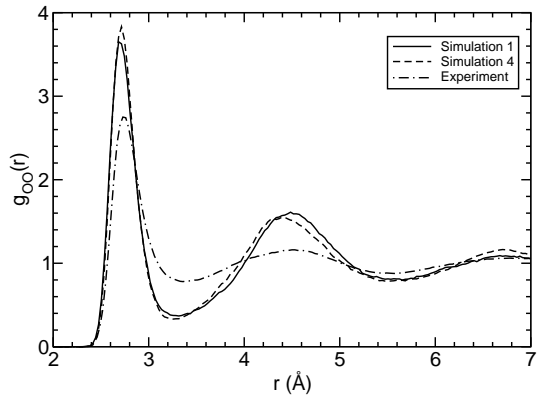}}}
\caption{Comparison of the oxygen-oxygen radial distribution functions
obtained in BO and CP simulations at a temperature of 300~K. The solid
line corresponds to CP dynamics, the dashed line to BO dynamics, and
the dot-dashed line to experiment \cite{asoper00b}.}
\label{gOOBOCP300}
\end{figure}

\begin{figure}
\rotatebox{-90}{\resizebox{2.5in}{!}{\includegraphics{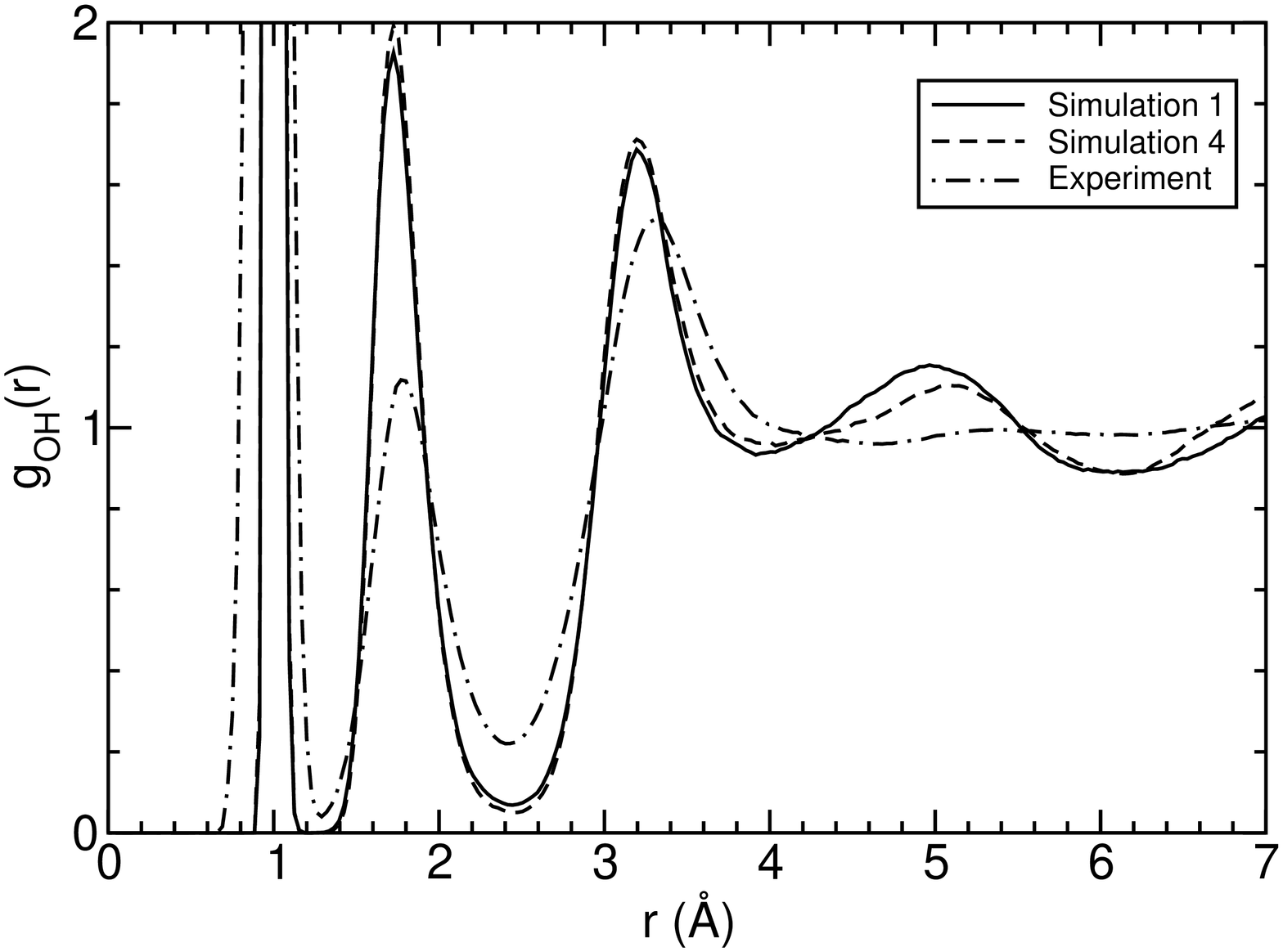}}}
\caption{Comparison of the oxygen-hydrogen radial distribution
functions obtained in BO and CP simulations at a temperature of
300~K. The solid line corresponds to CP dynamics, the dashed line to
BO dynamics, and the dot-dashed line to experiment \cite{asoper00b}.}
\label{gOHBOCP300}
\end{figure}

\begin{figure}[t]
\rotatebox{-90}{\resizebox{2.5in}{!}{\includegraphics{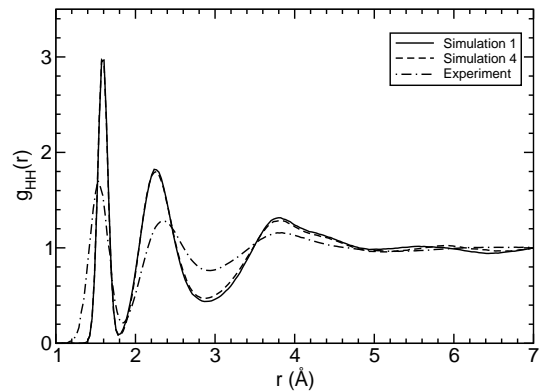}}}
\caption{Comparison of the hydrogen-hydrogen radial distribution
functions obtained in BO and CP simulations at a temperature of
300~K. The solid line corresponds to CP dynamics, the dashed line to
BO dynamics, and the dot-dashed line to experiment \cite{asoper00b}.}
\label{gHHBOCP300}
\end{figure}

In Figs.~\ref{gOOBOCP300} to \ref{gHHBOCP300} the oxygen-oxygen,
oxygen-hydrogen and hydrogen-hydrogen radial distribution functions
(RDFs) \cite{gofr} obtained with CP (Simulation 1) and 
BO (Simulation 4) {\it ab initio}
MD at a temperature of approximately 300~K are shown along with the
corresponding distribution functions obtained by experiment
\cite{asoper00b}.  As can be seen, both the BO and CP simulations of
water at 300~K lead to an overstructured liquid as compared to
experiment. In addition, these results confirm the previous
observation that, as long as $\mu$ is chosen appropriately, CP
simulations of water at ambient conditions can give a consistent
description of the liquid RDFs \cite{jgrossman04}.

The large amount of overstructure seen in Figs.~\ref{gOOBOCP300} to
\ref{gHHBOCP300} appears in the simulation within a relatively short
timescale ($\sim$2 ps), and can be readily observed, even when
starting the simulation from a classical theoretical model or a higher
simulation temperature, which may yield a less structured g(r). This
relatively short timescale should not be confused with the much longer
timescale required to obtain statistically independent data points for
a given RDF \cite{jgrossman04}, which will be discussed further in
Section \ref{timescales}.

In addition to the average structural properties of the liquid, we
have computed the diffusion coefficient using the Einstein relation:
\begin{equation}
6D = \lim_{t\rightarrow\infty}\frac{d}{dt}\left<\left|{\bf r}_i\left(t\right)
-{\bf r}_i\left(0\right)\right|^2\right>.
\label{einstein}
\end{equation}
To evaluate Eq.~\ref{einstein}, the mean square displacement (MSD) of
the oxygen atoms as a function of time was determined by averaging
over the trajectories with multiple starting configurations that are
evenly spaced by 4 fs. The slope of the resulting MSD was determined
in the range of 1 to 10 ps and used to compute the diffusion
coefficient, $D$.  As shown in Table \ref{summary}, the computed
diffusion coefficients obtained in this manner for both the CP and BO
simulations at 300~K are much smaller (by at least a factor of ten)
than D measured experimentally at the same temperature
\cite{rmills73}. Although the difference between the simulated and
measured diffusion coefficients are statistically meaningful, the
difference between the CP and BO diffusion coefficients at 300~K are
not. Indeed, to make a quantitative comparison between diffusion
coefficients that are in the range $10^{-7}$ to $10^{-6}cm^2/sec$,
orders of magnitude longer simulation times than those used here would
be required.

\onecolumngrid
\begin{widetext}
\begin{table}
\caption{For each simulation performed in this work, the type of
simulation (Car-Parrinello -- CP or Born-Oppenheimer -- BO, number of
molecules $N$, total simulation time (ps), average temperature (K),
diffusion coefficient D (cm$^2$/s), position (\AA) and value of first
maximum and minimum in the g$_{OO}$(r) \cite{gofr}, average temperature (K), and
average coordination (CN) of the water molecules are listed.  All
simulations are for H$_2$O with the PBE density functional. CP
simulations were carried out with a fictitious electron mass of 340
(1100) a.u. for flexible (rigid) water.  The last row contains
measured diffusion coefficients\cite{rmills73} and structural data
\cite{asoper00b} for water at ambient conditions.}
\begin{tabular}{ccccccccccc}
Sim. &
Type &
N &
Time  &
$T_{avg}$ &
$D$  &
R[g(r)$_{max}$] &
g(r)$_{max}$ &
R[g(r)$_{min}$] &
g(r)$_{min}$ &
CN \\ \hline
 1 & CP              &  54 &  19.8 & 296 & 2.4x10$^{-6}$ & 2.69 & 3.65 & 3.32 & 0.37 & 4.2 \\
 2 & CP              &  54 &  18.8 & 345 & 5.0x10$^{-6}$ & 2.72 & 3.21 & 3.35 & 0.42 & 4.3 \\
 3 & CP              &  54 &  22.0 & 399 & 2.2x10$^{-5}$ & 2.75 & 2.60 & 3.41 & 0.73 & 4.6 \\
 4 & BO              &  64 &  20.5 & 306 & 7.9x10$^{-7}$ & 2.72 & 3.83 & 3.25 & 0.33 & 4.1 \\
 5 & BO              &  64 &  20.7 & 349 & 3.3x10$^{-6}$ & 2.72 & 3.49 & 3.30 & 0.40 & 4.2 \\
 6 & BO              &  64 &  18.6 & 393 & 1.2x10$^{-5}$ & 2.73 & 3.10 & 3.40 & 0.56 & 4.6 \\
 7 & BO              &  64 &  10.5 & 442 & 3.9x10$^{-5}$ & 2.75 & 2.63 & 3.44 & 0.74 & 4.8 \\
 8 & CP-Rigid        &  54 &  24.5 & 315 & 1.3x10$^{-5}$ & 2.75 & 2.92 & 3.41 & 0.61 & 4.6 \\
 9 & CP-Rigid        &  54 &  26.3 & 345 & 3.3x10$^{-5}$ & 2.75 & 2.61 & 3.41 & 0.77 & 4.7 \\
10 & TIP5P(PIMC),P=1 &  64 & 500.0 & 300 & 2.6x10$^{-6}$ & 2.69 & 3.61 & 3.29 & 0.43 & 4.1 \\
11 & TIP5P(PIMC),P=1 &  64 & 500.0 & 350 & 2.8x10$^{-5}$ & 2.72 & 2.84 & 3.38 & 0.78 & 4.6 \\
12 & TIP5P(PIMC),P=5 & 216 &  ---  & 300 &      ---      & 2.73 & 2.76 & 3.44 & 0.77 & --- \\
   & Experiment      & --- &  ---  & 298 & 2.3x10$^{-5}$ & 2.73 & 2.75 & 3.36 & 0.78 & 4.7 \\
\hline
\end{tabular}
\label{summary}
\end{table}
\end{widetext}
\twocolumngrid

\subsection{Comparison of ab-initio MD to experiment and analysis of discrepancies}\label{errors}

There are a number of possible explanations for the observed
overstructure and slow diffusion of our DFT/GGA-based molecular
dynamics simulations of water.  In the following, we concentrate on
three of the most likely causes. The first is related to possible
inaccuracies of the PBE functional in describing hydrogen bonding. The
second is related to the neglect of proton quantum
effects. The third is related to the timescales 
used in the {\it ab initio} simulations.

\subsubsection{Accuracy of DFT/GGA for the water monomer and dimer}

In order to better quantify the accuracy of the DFT/GGA functional
used here, it is useful to first examine the computed properties of
the water monomer and dimer (for the interested reader, a more
comprehensive review of the water monomer and dimer properties as
computed within DFT was recently published in Ref.~\cite{xxu04}).

\begin{table}
\caption{Selected properties of the water monomer and dimer. All
reported distances are in \AA. DM refers to the dipole moment in
Debye, $\alpha_{iso}$ is the isotropic polarizability in \AA$^3$,
$OHO$ is the intermolecular hydrogen bond angle, and $D_e$ is the
binding energy of the water dimer in kcal/mol.}
\begin{ruledtabular}
\begin{tabular}{lccccccc}
 & \multicolumn{4}{c} {Monomer} & \multicolumn{3}{c} {Dimer}\\
\cline{2-5} \cline{6-8}
Method &
$r_{OH}$ &
$\angle HOH$ &
DM &
$\alpha_{iso}$ &
$r_{OO}$ &
$\angle OHO$ &
$D_e$ \\ \tableline
LDA     & 0.970     & 104.9     & 1.87     & 1.52     & 2.71     & 172.8     & 9.02 \\
PBE     & 0.970     & 104.2     & 1.81     & 1.55     & 2.90     & 173.7     & 5.10 \\
BLYP    & 0.972     & 104.5     & 1.80     & 1.54     & 2.95     & 171.6     & 4.18 \\
PBE1PBE & 0.959     & 104.9     & 1.85     & 1.42     & 2.90     & 174.1     & 4.90 \\
B3LYP   & 0.962     & 105.1     & 1.85     & 1.45     & 2.92     & 172.1     & 4.57 \\
CCSD(T) & 0.959$^a$ & 104.2$^a$ & ---      & ---      & 2.91$^b$ & 174.5$^b$ & 5.02$^b$ \\
Exp.    & 0.957$^c$ & 104.5$^c$ & 1.85$^d$ & 1.43$^e$ & 2.98$^f$ & 174.0$^f$ & 5.44$^g$ \\
\end{tabular}
\footnotetext{
$^a$Ref.~\cite{gtschumper02}.
$^b$Ref.~\cite{wklopper00}.
$^c$Ref.~\cite{wbenedict75}.
$^d$Ref.~\cite{sclough73}.
$^e$Ref.~\cite{wmurphy77}.
$^f$Ref.~\cite{jodutola80}.
$^g$Ref.~\cite{lcurtiss79}.}
\end{ruledtabular}
\label{accuracy}
\end{table} 

As shown in Table \ref{accuracy}, we have examined the use of a number
of DFT exchange-correlation functionals, including LDA
\cite{jslater51,svosko88}, GGA (PBE \cite{jperdew96} and BLYP
\cite{abecke88,clee88}), and hybrid (PBE1PBE \cite{mernzerhof99} and
B3LYP \cite{abecke93}) functionals. Also included in Table
\ref{accuracy} are the results of several CCSD(T) calculations
\cite{wklopper00,gtschumper02} as well as the relevant experimental
measurements
\cite{wbenedict75,sclough73,wmurphy77,jodutola80,lcurtiss79}.  The
CCSD(T) results represent the most accurate quantum chemistry
computations of the water monomer and dimer to date. For all of the
DFT-based calculations reported in Table \ref{accuracy} we have used
the electronic structure program Gaussian 03 \cite{g03} with the
aug-cc-pVTZ basis set \cite{rkendall92}.

For the computed bond distances and angles, the majority of the tested
functionals perform reasonably well. The most notable exception is the
oxygen-oxygen distance in the LDA water dimer calculation, which is
much too short. Although the computed dipole moments of the water
monomer are also in good agreement with each other, it is interesting
to note that the isotropic polarizabilities are systematically too
large at the LDA and GGA level of theory. This is due to the
well-known tendency of local DFT functionals to underestimate the
HOMO-LUMO gap of molecular systems, which in turn leads to an
overestimation of the molecular polarizability. As more accurate,
hybrid functionals are considered, this general tendency seems to
disappear.  In Table \ref{accuracy} we also show computed binding
energies. As expected, the LDA binding energy is much too large. The
remainder of the DFT functional binding energies are in relatively
good agreement with each other, as well as with the CCSD(T)
calculation and the experimental measurement.

Given the differences in the water dimer binding energies found with
BLYP and PBE (the BLYP binding energy is $\sim$1 kcal/mol smaller than
PBE), one might think that the PBE functional would give more
structured RDFs and slower diffusion of the liquid, as compared to
BLYP.  In our previous work we could not resolve any statistically
significant difference between PBE and BLYP results. However,
comparison of the present CP results with those of Chen {\it et al}.
\cite{bchen02} -- which were obtained with a sufficiently small
fictitious mass ($\mu$=400 a.u.) and the BLYP functional -- seem to
show differences which would be consistent with the idea that BLYP
yields less structured RDFs than PBE. At this point, it is difficult
to draw any firm conclusions, as the time scale of both the present
work and the simulations of Ref.~\cite{bchen02} are probably too small
to statistically resolve differences. 

We note that there is a very good agreement between the PBE and the
CCSD(T) binding energies, and overall the water monomer and dimer
properties listed in Table \ref{accuracy} appear to be well reproduced
by the PBE functional. Although these zero temperature, gas phase results do not
necessarily insure an equally good performance of the functional in
the liquid state, they are very promising.  We also note that the PBE
functional appears to be quite accurate in reproducing the sublimation
energy, lattice constant and bulk modulus of ice-Ih
\cite{dhamann97}. It would be interesting to explore the accuracy of
other functionals, along the lines proposed in
Ref.\cite{yzang98,dasthagiri03}, especially of hybrid functionals, but
this falls beyond the scope of the present investigation.

\subsubsection{Proton quantum effects}\label{qeffects}

In addition to possible inaccuracies in the DFT/PBE treatment of
water, it is plausible that some amount of the observed overstructure
is due to the neglect of the quantum motion of the hydrogen atoms.  As
pointed out in Ref.~\cite{itironi96}, at 300~K, $k_BT\sim200$
cm$^{-1}$, whereas the high frequency intramolecular modes in water
range from $\sim$1000 to 3500 cm$^{-1}$. Therefore, in the real
quantum system, the amount of thermal energy available to excite
vibrational modes is much smaller than the lowest possible
intramolecular vibrational excitations,
\begin{equation}
\hbar\omega \gg k_BT.
\end{equation}
In other words, the quantum system will be restricted to its
vibrational ground state at a temperature of 300~K, and the quantum
and the classical system will be qualitatively different in terms of
the distribution of vibrational energy. When using empirical
potentials with intramolecular flexibility, quantum effects can, to
some extent, be implicitly included in the potential
parameterization. However, in the case of {\it ab initio} based
methods, the possibility of implicitly accounting for quantum effects
by renormalizing the interaction potential is, by definition, not an
option.

\begin{figure}
\rotatebox{-90}{\resizebox{2.5in}{!}{\includegraphics{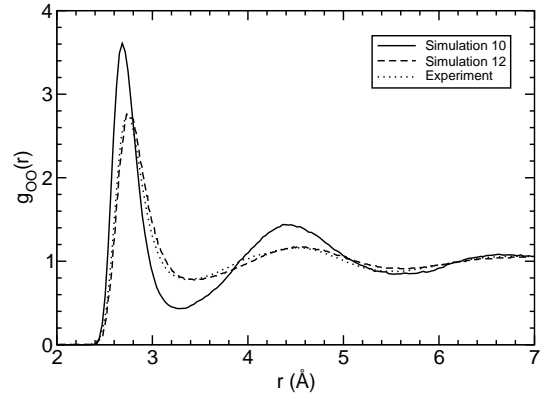}}}
\caption{The oxygen-oxygen radial distribution function obtained in
TIP5P(PIMC) simulations (see text) of water at a density
of 0.997 g/cc, and a temperature of 300 K. The solid line corresponds
to a simulation performed with P=1 (no path integral sampling), the
dashed line to P=5 (taken from Ref.~\cite{mmahoney01}), and the dotted
line to experiment \cite{asoper00b}.}
\label{TIP5PPIMC}
\end{figure}

In principle, the effect of the proton quantum motion can be accounted
for with path-integral (PI) methods \cite{rfeynman65}. To date,
combining PI sampling of ionic trajectories with empirical interaction
potentials has been found to lead to an overall softening of the RDFs
of water at ambient conditions
\cite{rkuharski85,gbuono91,jlobaugh97,hstern01,mmahoney01}. However,
the majority of these simulations have been performed with empirical
potentials that already include some amount of quantum effects.  In
particular, it is common to use classical molecular dynamics or Monte
Carlo sampling in the parameterization of an interatomic potential, with
the goal of fitting experimental data as well as possible.  Within
this procedure, agreement with experiment is achieved by renormalizing
interatomic interactions so as to include proton quantum effects,
which are not explicitly treated by classical molecular dynamics or
Monte Carlo.  There are, however, important exceptions to this general
procedure. In particular, in Ref.~\cite{mmahoney01}, the TIP5P
potential was reparameterized with PI Monte Carlo sampling instead of
classical Monte Carlo sampling. The resulting potential, which we will
refer to here as TIP5P(PIMC), is nearly identical to the original
TIP5P potential except for the magnitude of the charges on the
hydrogen and lone-pair sites, which have been slightly increased (see
Table \ref{watermodels}). The TIP5P(PIMC) potential opens up the
interesting possibility of directly examining, in the context of a
rigid water model, the extent to which quantum effects can soften the
RDF of water. As shown in Fig.~\ref{TIP5PPIMC}, when the TIP5P(PIMC)
potential is used in a classical molecular dynamics simulation with
P=1 (P refers to the number of beads used to represent the path
integrals), the resulting oxygen-oxygen RDF becomes severely
overstructured as compared to a PI Monte Carlo calculation with the
TIP5P(PIMC) potential where P=5 is used (taken from
Ref.~\cite{mmahoney01}). The amount of overstructure seen in
Fig.~\ref{TIP5PPIMC} is nearly identical to the BO and CP RDFs
presented earlier in Fig.~\ref{gOOBOCP300}.  In addition, in the
TIP5P(PIMC), P=1 simulation the diffusion coefficient of the liquid is
$2.2\times10^{-6}cm^2/s$, which is approximately ten times smaller
than the experimental value; this value is also in qualitative
agreement with the corresponding BO and CP simulations performed at
300~K (see Table \ref{summary}). These results indicate that including
proton quantum effects may bring the {\it ab initio} results presented here
into significantly better agreement with experiment and that neglect of
these effects is probably responsible for the majority of the
discrepancy between experiment and DFT results.

\begin{figure}
\rotatebox{-90}{\resizebox{2.5in}{!}{\includegraphics{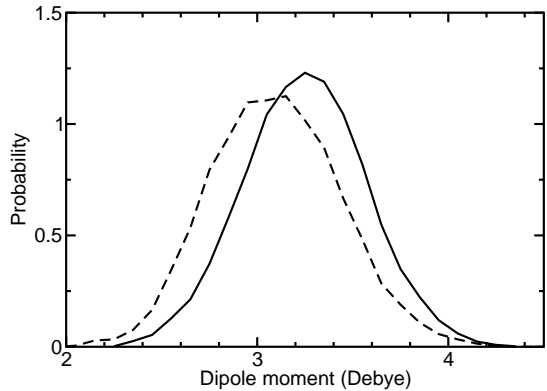}}}
\caption{The probability distribution of water molecule dipole moments
obtained in CP simulations of water at a temperature of 300 K. The
solid line corresponds to a simulation performed with the fictitious
mass $\mu$=340 au, and the dashed line to a simulation with $\mu$=760
au.}
\label{Dipoles}
\end{figure}

Along with the existing PI studies based on empirical water potentials
\cite{rkuharski85,gbuono91,jlobaugh97,hstern01,mmahoney01}, Chen, {\it
et al.}\cite{bchen03} recently demonstrated the feasibility of
directly performing PI sampling and CP simulations of water. However,
instead of the expected softening of the liquid structure, Chen, {et
al.}\cite{bchen03} found the rather surprising result that quantum
effects may actually {\it increase} the structure of water. The reason
for this increase was attributed to an enhancement of the individual
water molecule's dipole moments when proton quantum effects are
included in the calculations.  In particular, an analysis based on
maximally localized Wannier functions \cite{psilvestrelli99} was used
to demonstrate that the average dipole moment of water molecules
increases from 3.08 Debye in their CP simulation performed with
classical ion dynamics to an average value of 3.19 Debye when PI
sampling is included. Although we have used a different GGA functional
(PBE instead of the BLYP functional used in Ref.~\cite{bchen03}), it
is interesting to note that in our CP simulations we find a dipole
moment of 3.24 Debye when a fictitious mass of $\mu$=320 au
is used (a value of $\mu$=500 au was reported in 
Ref.~\cite{bchen03}).  As shown in Fig.~\ref{Dipoles}, when a larger
value of the ficticious mass is used ($\mu$=740 au) the average dipole
moment decreases to 3.06 Debye.  An average value of 3.24 Debye found
here is not consistent with the results reported in
Ref.~\cite{bchen03}, and opens up the possibility that the role of
quantum effects in path integral/DFT-based simulations of water may
not be completely understood. Indeed, simulations using PI sampling
with a flexible water model are known to be quite difficult to
converge \cite{mmahoney01}.  In this regard, additional
investigations, which are beyond the scope of this work, are most
likely needed where the convergence of the various technical
parameters involved in PI simulations are examined in detail.

\subsubsection{Timescales}\label{timescales}
\begin{figure}
\rotatebox{-90}{\resizebox{2.5in}{!}{\includegraphics{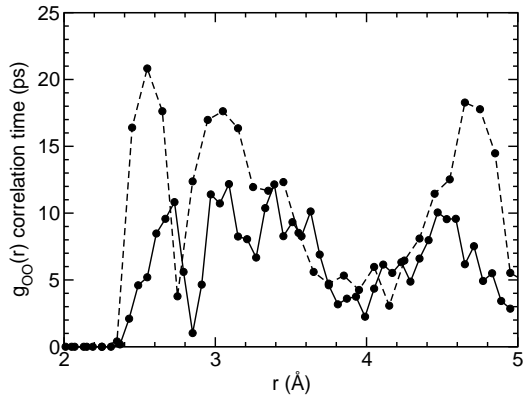}}}
\caption{
The oxygen-oxygen RDF correlation
times computed from classical molecular dynamics simulations using
the TIP5P (solid line) and TIP5P(PIMC), P=1 (dashed line) potentials
at ambient conditions. The points correspond to the
correlation time for each bin that was used to collect the
distribution function.}
\label{correlation}
\end{figure}

As discussed in Ref.~\cite{jgrossman04}, another approximation that needs
to be considered is the length of the time interval over which averages,
such as the RDFs, are taken. In the case of {\it ab initio}
molecular dynamics simulations, which are usually limited to relatively
short timescales (10 to 20 ps), this can be a particularly significant
effect. In this regard, it is important to determine the time interval
outside which neighboring simulation data points are no longer correlated,
and thus statistically significant. In Ref.~\cite{jgrossman04} a
method based on a combination of re-blocking and autocorrelation
function techniques was used for determining this time interval along
with an estimate of the variance for individual bins of a RDF. Because this
method requires simulation trajectories of at least 1 to 2 orders of
magnitude longer than the actual correlation time,
a 2 ns classical MD
simulation of 32 TIP5P water molecules under ambient conditions was
used for the analysis. Based on this trajectory, correlation times as long
as 10 ps were observed for the oxygen-oxygen RDF. However, we note 
that the diffusion coefficient for 32 TIP5P water molecules under
ambient
conditions is $\sim 2.1\times10^{-5}cm^2/s$, which is at least 10 times larger
than what is found in our {\it ab initio}
BO and CP simulations at 300~K. Therefore, in
order
to have an estimate of the RDF correlation time and variance 
relevant to our 300~K BO and CP simulations, it is more appropriate to use the
TIP5P(PIMC), P=1 potential for this analysis. In Fig.~\ref{correlation},
the computed correlation time for the oxygen-oxygen RDF bins is shown for
both
TIP5P and the TIP5P(PIMC), P=1 simulations of ambient water. As can be seen
near the regions of the first peak in $g_{OO}(r)$, the correlation times
increase from $\sim10$ ps with the TIP5P potential to $\sim20$ ps for the
TIP5P(PIMC) potential. Although there is a considerable amount of scatter 
in the correlation times shown in Fig.~\ref{correlation}, 
this analysis does provide a useful lower bound 
to the expected correlation times of the radial distribution 
function bins. 

\begin{figure}
\rotatebox{-90}{\resizebox{2.5in}{!}{\includegraphics{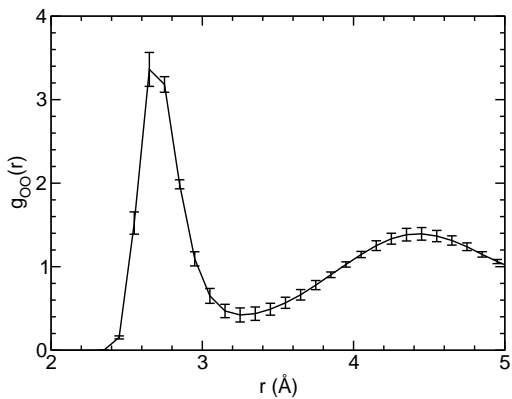}}}
\caption{Oxygen-oxygen RDF for the TIP5P(PIMC), P=1 classical potential.
The error bars correspond to the square root of the variance.}
\label{variance}
\end{figure}

In Fig.~\ref{variance}, an estimate of the variance in the
oxygen-oxygen RDF bins is shown for a block of length equal to the
TIP5P(PIMC)
correlation times displayed in Fig.~\ref{correlation}. As discussed in
Ref.~\cite{jgrossman04}, the error bars shown in Fig.~\ref{variance} are
representative of the uncertainties that should be expected when comparing
independent $\sim20$ ps simulations of water that are characterized by
a diffusion coefficient on the order of $10^{-6}cm^2/s$.

\begin{figure}
\rotatebox{-90}{\resizebox{2.5in}{!}{\includegraphics{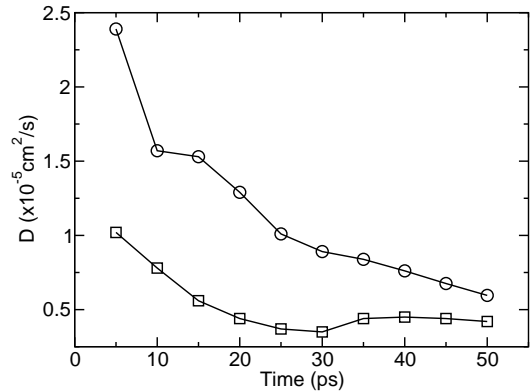}}}
\caption{The computed diffusion coefficient as a function of simulation
time for the TIP5P(PIMC), P=1 potential at a temperature of
300 K and a density of 0.997 g/cc. The circles correspond to a simulation
that was initially equilibrated with the TIP5P potential at 600 K and
the squares to a simulation that was equilibrated with the TIP5P
potential at 300 K (see text for details).}
\label{diff}
\end{figure}

In addition to uncertainties in structural properties, the limited
timescales of a typical first-principles simulation of water
can strongly influence dynamical properties such as computed
diffusion coefficients. In order to illustrate this effect we have
performed two simulations of 64 water molecules with the TIP5P(PIMC), P=1
potential that differ in how the starting configurations were
equilibrated. In the first simulation, the initial starting configuration
was taken from a 500 ps TIP5P simulation at a temperature of 300 K. The
simulation was then continued for an additional 53 ps with the TIP5P(PIMC),
P=1 potential at a temperature of 300 K. This simulation closely resembles 
the equilibration procedure that was used for the first-principles 
simulations presented here. In the second
simulation, the initial starting configuration was taken from a 500 ps
TIP5P simulation at a temperature of 600 K. The simulation was then
continued, as before, for an additional 53 ps with the TIP5P(PIMC), P=1
potential at a temperature of 300 K. The last 50 ps (the first 3 ps were
discarded) of these two simulations were analyzed by computing the
diffusion coefficient via Eq.~\ref{einstein} as a function of the simulation
time at 5 ps intervals. As can be seen in Fig.~\ref{diff}, the specifics of
the initial equilibration phase on the diffusion coefficients
can be rather dramatic. In particular, an equilibration phase that involves 
annealing at a higher temperature can lead to a significant overestimation 
of the diffusion coefficient with respect to the infinite time limit, 
even for simulations as long as $\sim 50 ps$.

\subsection{The effect of temperature in simulations of water}\label{temp}

\begin{figure}
\rotatebox{-90}{\resizebox{2.5in}{!}{\includegraphics{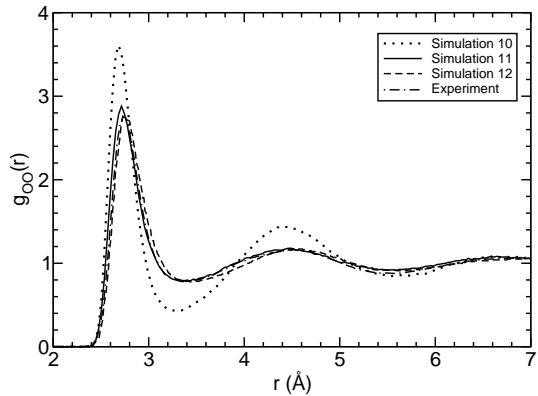}}}
\caption{The oxygen-oxygen radial distribution function obtained in
different TIP5P(PIMC) simulations. The dotted line corresponds to a
simulation performed at 300 K with P=1, the solid line to 350 K with
P=1, the dashed line to 300 K with P=5 (taken from
Ref.~\cite{mmahoney01}), and the dashed-dotted line to experiment
\cite{asoper00b}.}
\label{TIP5PPIMCP1P5}
\end{figure}

As pointed out in Ref.~\cite{rkuharski85} for the ST2 water model, and
in Ref.~\cite{gbuono91} for the SPC and TIP4P models, the main
structural changes that occur when going from a classical to a quantum
description of water at 300~K are similar to simply increasing the
temperature of the classical simulation by 50~K.  In order to examine
if similar results are obtained when the TIP5P(PIMC) potential is
used, we have performed classical simulations (P=1) of water with the
TIP5P(PIMC) potential at temperatures of 300 and 350~K (simulations 10
and 11 in Table \ref{summary}).  In Fig.~\ref{TIP5PPIMCP1P5}, the
oxygen-oxygen RDF obtained in these simulations are compared to
TIP5P(PIMC), P=5 at 300~K \cite{mmahoney01}, and to experiment at
300~K \cite{asoper00b}. As expected from previous investigations
\cite{rkuharski85,gbuono91}, the main structural differences between
the classical and quantum fluid described by the TIP5P(PIMC) are
remarkably similar to a 50~K temperature increase in the classical
simulation temperature at constant density. In addition to the observed 
structural
changes, the 50~K temperature increase also causes the diffusion
coefficient of classical TIP5P(PIMC) water to increase from
$2.2\times10^{-6}cm^2/s$ to $3.0\times10^{-5}cm^2/s$, which is in much
better agreement with the experimental measurement of
$2.2\times10^{-5}cm^2/s$ for water under ambient conditions
\cite{rmills73}. We also note that approximately accounting for
quantum effects via temperature scaling is a technique that has been
used in a variety of materials other than water
\cite{cwang90,lporter97}.

\begin{figure}
\rotatebox{-90}{\resizebox{2.5in}{!}{\includegraphics{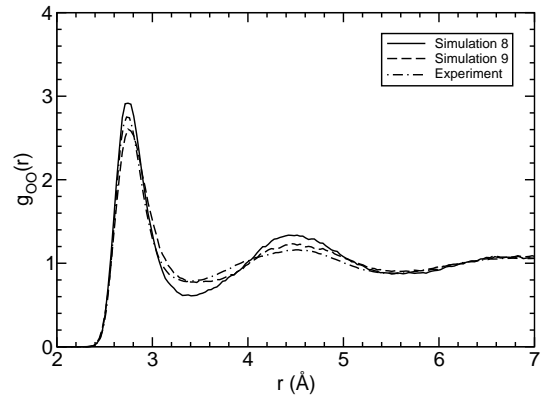}}}
\caption{The oxygen-oxygen radial distribution function obtained in CP
simulations of rigid water. The solid line corresponds to a simulation
performed at 315 K, the dashed line to a simulation at 345 K, and the
dashed-dotted line to experiment \cite{asoper00b}.}
\label{RigidT}
\end{figure}

Given the agreement of results obtained with the TIP5P(PIMC) potential
when including quantum effects, and increasing the classical
simulation temperature, it is interesting to examine if a similar
decrease in structure occurs in DFT simulations of water as a function
of temperature. However, it is important to note that the TIP5P(PIMC)
potential used above is based on a rigid water model that does not
allow for intermolecular flexibility.  Therefore, to make a direct
comparison between classical and {\it ab initio} MD, we have first examined
the effects of increasing the temperature in CP simulations of rigid
water. Our results are reported in Fig.~\ref{RigidT}. The technical
details of the simulations are the same as those used in simulation B
of Ref.~\cite{mallesch04}.  As can be seen by comparing
Figs.~\ref{gOOBOCP300} and \ref{RigidT}, although the amount of
structure in the CP rigid water simulation at a temperature of 315 K
(simulation 8) is smaller than the one observed in the flexible water
simulation (simulation 1 at 296~K), there is still a noticeable amount of
overstructure as compared to experiment. Also shown in
Fig.~\ref{RigidT} are the results of a CP simulation of rigid water at
a temperature of 345~K (simulation 9). As was seen in the case of the
TIP5P(PIMC) potential, at a temperature of $\sim$350~K the
oxygen-oxygen radial distribution function is in good agreement with
the corresponding experimental measurement at 300~K. Therefore, we
expect that an increase in temperature has a similar effect on the
liquid structure with both the TIP5P(PIMC) potential and DFT/rigid
water. This provides indirect evidence that RDFs computed with
TIP5P(PIMC) and DFT/rigid water will soften to the same extent when PI
sampling is performed.

\begin{figure}
\rotatebox{-90}{\resizebox{2.5in}{!}{\includegraphics{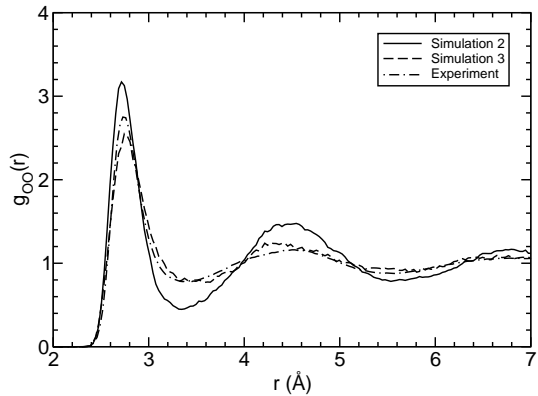}}}
\caption{The oxygen-oxygen radial distribution function obtained in CP
simulations of water at different temperatures. The solid line
corresponds to a simulation performed at $\sim$350 K, the dashed
line to $\sim$400 K, and the dashed-dotted line to experiment \cite{asoper00b}.}
\label{CPMDTOO}
\end{figure}

\begin{figure}
\rotatebox{-90}{\resizebox{2.5in}{!}{\includegraphics{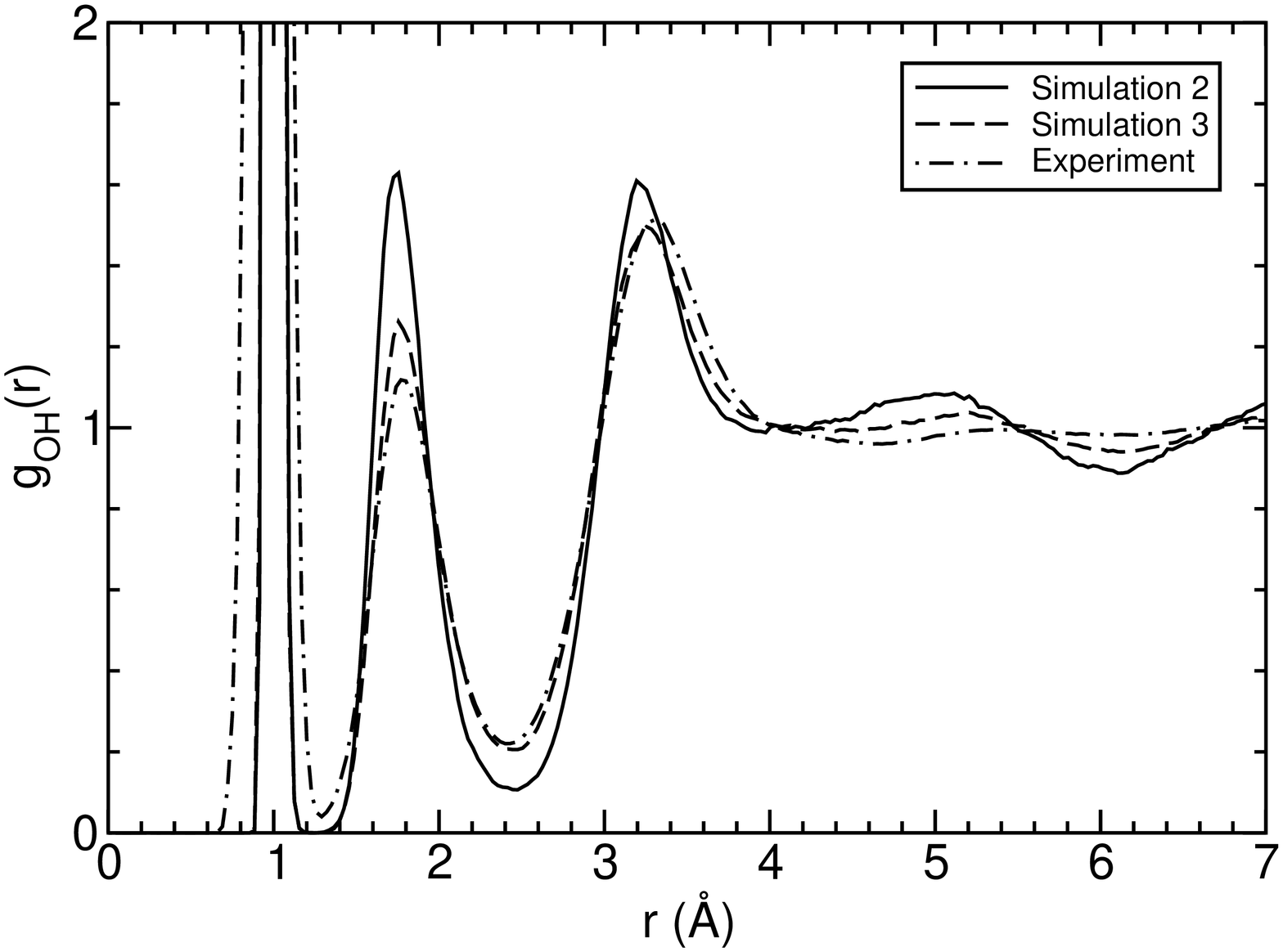}}}
\caption{The oxygen-hydrogen radial distribution function obtained in
CP simulations of water at different temperatures. The solid line
corresponds to a simulation performed at $\sim$350 K, the dashed line
to $\sim$400 K, and the dashed-dotted line to experiment \cite{asoper00b}.}
\label{CPMDTOH}
\end{figure}

\begin{figure}
\rotatebox{-90}{\resizebox{2.5in}{!}{\includegraphics{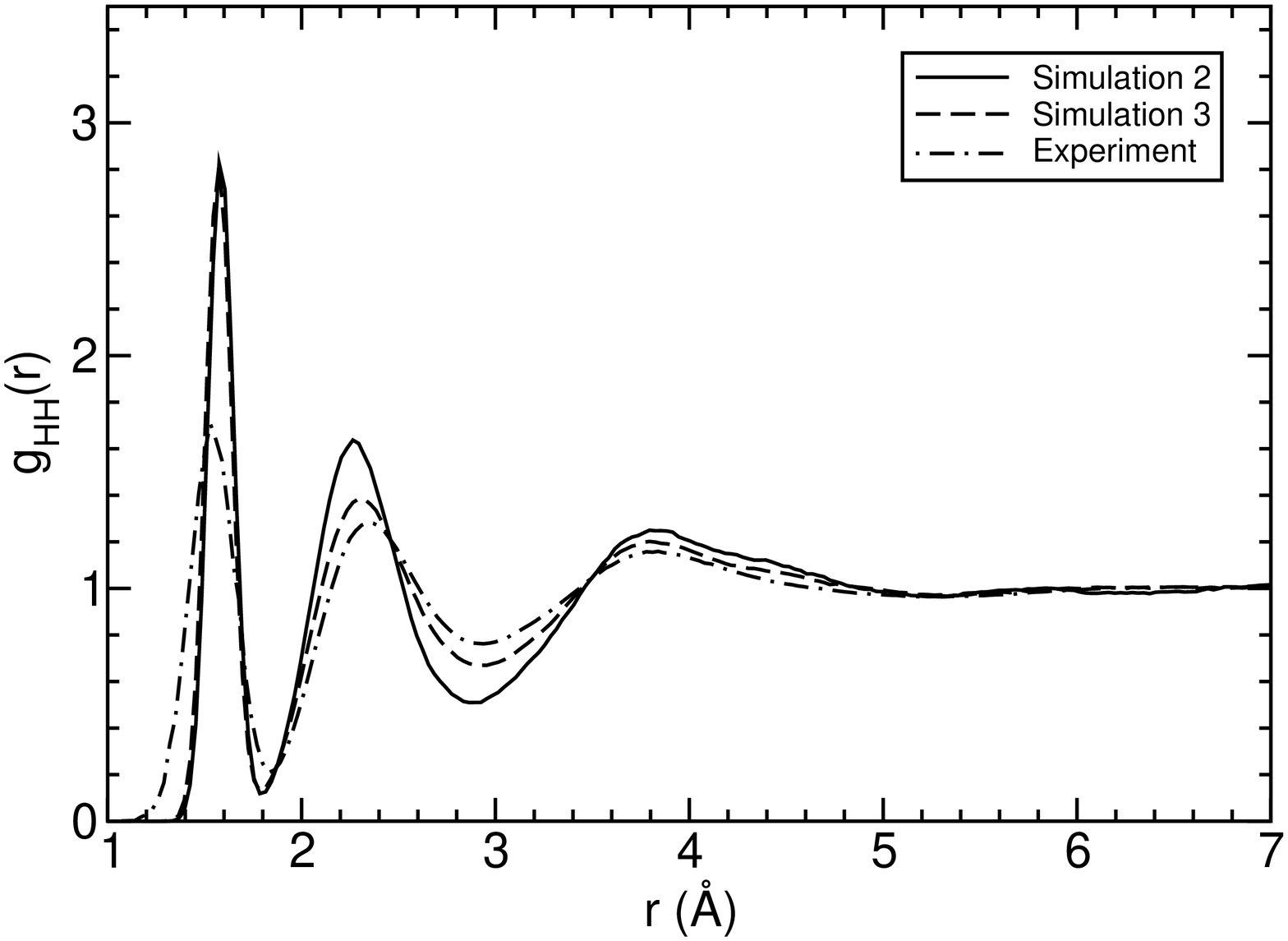}}}
\caption{The hydrogen-hydrogen radial distribution function obtained
in CP simulations of water at different temperatures. The solid line
corresponds to a simulation performed at $\sim$350 K, the dashed line
to $\sim$400 K, and the dashed-dotted line to experiment \cite{asoper00b}.}
\label{CPMDTHH}
\end{figure}

\begin{figure}
\rotatebox{-90}{\resizebox{2.5in}{!}{\includegraphics{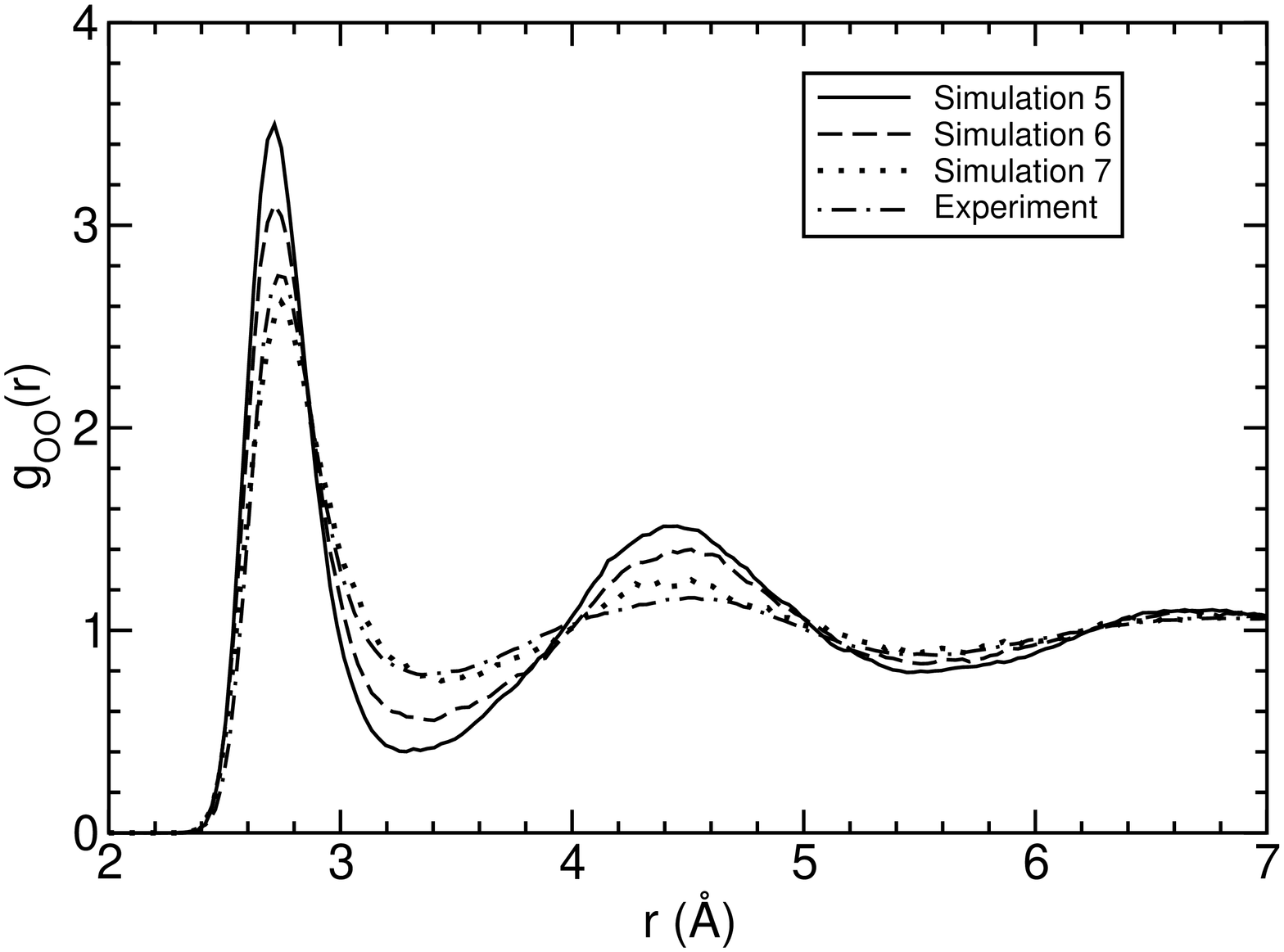}}}
\caption{The oxygen-oxygen radial distribution function obtained in BO
simulations of water at different temperatures. The solid line
corresponds to a simulation performed at $\sim$350 K, the dashed line
to $\sim$400 K, the dotted line to $\sim$450 K, and the dashed-dotted
line to experiment \cite{asoper00b}.}
\label{BOMDTOO}
\end{figure}

\begin{figure}
\rotatebox{-90}{\resizebox{2.5in}{!}{\includegraphics{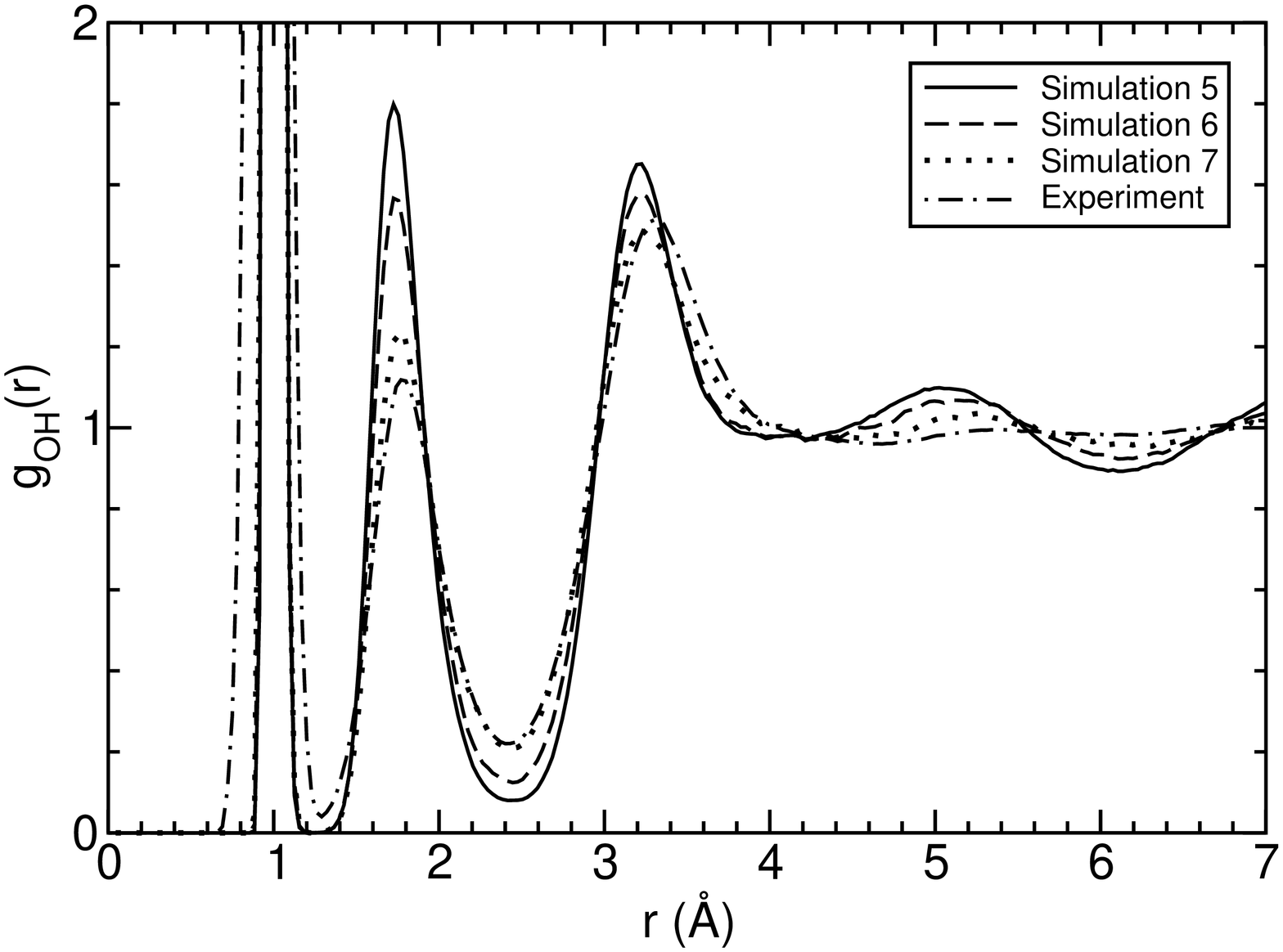}}}
\caption{The oxygen-hydrogen radial distribution function obtained in
BO simulations of water at different temperatures. The solid line
corresponds to a simulation performed at $\sim$350 K, the dashed line
to $\sim$400 K, the dotted line to $\sim$450 K, and the dashed-dotted
line to experiment \cite{asoper00b}.}
\label{BOMDTOH}
\end{figure}

\begin{figure}
\rotatebox{-90}{\resizebox{2.5in}{!}{\includegraphics{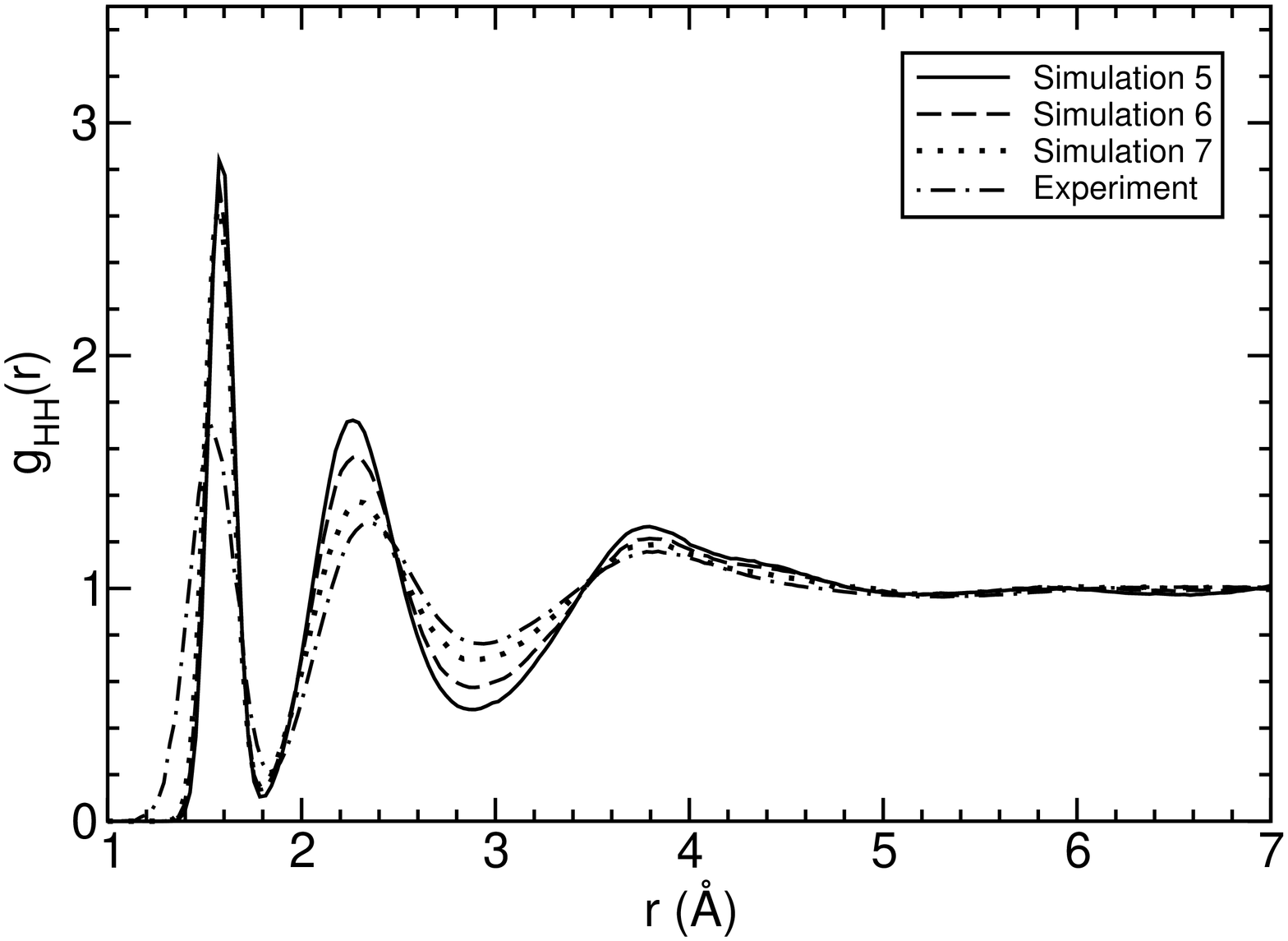}}}
\caption{The hydrogen-hydrogen radial distribution function obtained
in BO simulations of water at different temperatures. The solid line
corresponds to a simulation performed at $\sim$350 K, the dashed line
to $\sim$400 K, the dotted line to $\sim$450 K, and the dashed-dotted
line to experiment
\cite{asoper00b}.}
\label{BOMDTHH}
\end{figure}

In addition to CP simulations of rigid water, we have also
investigated the effect of temperature in CP and BO simulations of
flexible water at constant density. As shown in Figs.~\ref{CPMDTOO} 
to \ref{CPMDTHH} for
CP dynamics, and in Figs.~\ref{BOMDTOO} to \ref{BOMDTHH} for BO
dynamics, an increase in the simulation temperature decreases the
overstructure in the radial distribution functions found in the
corresponding simulations at 300~K. However, close inspection of
Figs.~\ref{CPMDTOO} to \ref{BOMDTHH} reveals that both CP and BO
simulations of flexible water require a significantly larger
temperature increase than the simulations of rigid water in order to
bring the radial distribution functions into agreement with experiment
at 300~K.

\begin{figure}
\rotatebox{-90}{\resizebox{2.5in}{!}{\includegraphics{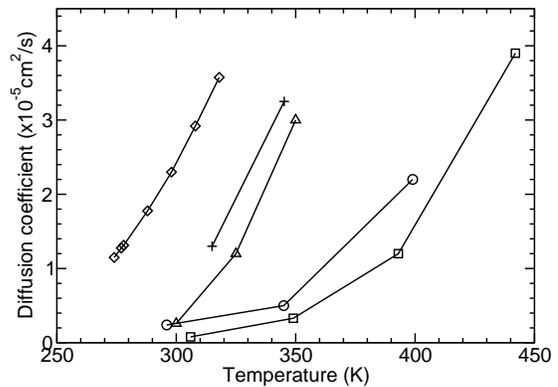}}}
\caption{The diffusion coefficient of water as a function of
temperature.  The diamonds correspond to the experimental measurements
reported in Ref.~\cite{rmills73}, crosses to rigid water CP, triangles
to TIP5P(PIMC), P=1, circles to flexible water CP, and squares to
flexible water BO. The lines are simply a guide to the eye.}
\label{diffusion}
\end{figure}

In Fig.~\ref{diffusion} the computed diffusion coefficients for
different simulations are shown as a function of the simulation
temperature.  Also shown in Fig.~\ref{diffusion} are the
experimentally measured diffusion coefficients of water as a function
of temperature \cite{rmills73,kkrynicki78}. Consistent with the
oxygen-oxygen radial distribution functions shown in
Fig.~\ref{RigidT}, the temperature dependence of the CP rigid water
diffusion coefficient is similar to that of TIP5P(PIMC), P=1. This
indicates that at approximately 350~K the simulation diffusion
coefficient obtained with CP-rigid water simulations is equal to the
experimental value at 300~K.

In the flexible water simulations, the temperature dependence of the
diffusion coefficients are quite different. Over the same temperature
range of 300 to 350~K, the computed diffusion coefficients increase at
a much slower rate than in the rigid water simulations, and
temperatures of 400~K in the CP simulation and $\sim$ 415~K in the BO
simulations are needed in order to obtain diffusion coefficients
comparable to experiment at 300~K. These findings are in qualitative
agreement with previous observations based on empirical potentials,
which found that the use of a rigid water approximation tends to lead
to an overall decrease in structure of the liquid and faster diffusion
as compared to a flexible representation with the same intramolecular
potential \cite{jlobaugh97}.

We note that for all of the temperatures considered here, the
diffusion coefficients obtained with CP dynamics (with $\mu$=340 a.u.)
appear to be approximately two times larger than in the corresponding
BO simulations. As mentioned earlier in Section \ref{BOvsCP} and
discussed in more detail by Tangney {\it et al.}  \cite{ptangney02b},
this difference between the computed diffusion coefficients in CP and
BO simulations is most likely due to a small, but still significant,
effect of the chosen ficticious mass value in the CP simulations.

\section{Conclusion}\label{conc}

We have presented a series of 20 ps {\it ab initio} MD simulations 
of water
at ambient conditions and temperatures ranging from 300 to 450~K,
carried out using both the CP and BO techniques. Radial
distribution functions obtained with both approaches are in excellent
agreement with each other, provided an appropriately small fictitious
mass is used in CP simulations. Differences have been observed in
computed diffusion coefficients, those obtained with BO dynamics being
systematically $\sim$ 2 times smaller than the corresponding CP
values. However, these differences are statistically significant only
at T $\simeq$ 400~K, given the time span of our simulations (20
ps). These results indicate that even for relatively small values of
the fictitious electronic mass, some coupling between the classical
motion of electronic and ionic degrees of freedom is unavoidable in CP
simulations, as pointed out by Tangney {\it et al.}\cite{ptangney02b}. 
Furthermore, the inaccuracies
introduced by this coupling appear not to have a significant,
quantitative effect on calculated structural properties; rather, their
effect may be seen on dynamical properties such as the diffusion
coefficient.

We have also presented an analysis of the effects of the proton
quantum motion on computed structural and dynamical properties of
water. Our results, obtained with a series of classical MD and PI
calculations with empirical force fields, provide strong indications
that the neglect of quantum effects may account for most of the
discrepancy observed here between DFT-based simulations and
experiment. We have found that including proton quantum effects is
approximately equivalent to a temperature increase of 50 and 100~K,
when using a rigid or a flexible water model, respectively, under 
constant volume conditions. This is true for both RDFs and diffusion 
coefficients. We note that although quantum effects most certainly do have an 
influence on the properties of water at ambient conditions, the 
exact magnitude of this effect is still unclear. For example, all 
of the simulations shown here have been performed under constant 
volume conditions that corresponds to a density of $0.997 g/cc$. 
It is likely that this does not correspond to the exact equilibrium 
density of water as described by the PBE functional, which may in 
turn have an effect on the precise temperatures needed to bring the 
RDFs and diffusion coefficients into agreement with experiment. 

Finally, we note that inaccuracies in the DFT description of hydrogen
bonding may be attributed to the choice of the local energy functional
(PBE in this paper); the use of hybrid functionals that provide a
better agreement between measured and calculated gas phase
polarizabilties may improve the description of the fluid as well.

\section{Acknowledgments}
We thank L.R.~Pratt, D.~Asthagiri and D. Prendergast for many useful 
discussions, and E.W.~Draeger for performing the variance calculations 
used here.  
This work was performed under the auspices of the
U.S.~Dept.~of Energy at the University of California/Lawrence
Livermore National Laboratory under contract no.~W-7405-Eng-48.


\begin{thebibliography}{10}
\bibitem{gofr}
All radial distribution functions computed between identical atom types have
  been normalized by $\Omega/N\left(N-1\right)$ where $\Omega$ is the volume of
  the simulation cell and $N$ is the number of atoms involved in the
  distribution function.
\end{thebibliography}
\end{document}